\documentclass[]{article}

\title{Steady linked vortices with chaotic streamlines}

\author{Oscar Velasco Fuentes\\
            Departamento de Oceanograf\'{\i}a F\'{\i}sica\\
            CICESE, Ensenada, M\'exico,
            ovelasco@cicese.mx      
}
\date{13 October 2010}

\begin{document}

\maketitle

\begin{abstract}
This paper contains background information for a fluid dynamics video
submitted to the Gallery of Fluid Motion to be held along with the  63rd Annual Meeting 
of the American Physical Society's Division of Fluid Dynamics.
\end{abstract}

After observing the smoke-ring demonstrations of P.G. Tait in January 1867, 
Kelvin  wrote an enthusiastic letter to Helmholtz where he put forward his
hypothesis of the {\em{vortex atom}} and outlined some of the vortex problems
that would occupy him in the following years. This is what he wrote about linked vortices:
``I am, as yet, a good deal puzzled as to what two vortex-rings through one another would do 
(how each would move, and how its shape would be influenced by the other)'' [1].

Later Kelvin [3] and J.J. Thomson [2] would describe the conditions under which
linked vortices could be both steady and stable.
Following them, we assume that two filamentary vortices of equal strength 
lie on the surface of an immaterial torus. 
Each vortex is uniformly coiled on the torus in such a way that, before closing on itself, 
it winds once around the torus' symmetry axis and once around the torus' centreline.
Furthermore, the vortices intersect any meridional plane
at opposite extremes of a diameter of the torus' cross section.
We represent each ring vortex with a finite number of material markers,
compute their velocities with the Rosenhead-Moore approximation to the 
Biot-Savart law, and move them with a fourth-order Runge-Kutta scheme
(see [4]).

The first part of the video shows the motion of the vortices (red and blue tubes)
as well as of a set of passive tracers (white dots) in a fixed frame of reference.
The vortices progress along and rotate around the torus' symmetry axis
in an almost steady manner while preserving their shape.
The second part shows the initial and final periods of the same simulation 
but this time viewed in the comoving frame of reference. 
Note that while the vortices remain stationary 
the fluid particles that make them up move along the almost permanent shapes 
(the black strips, as well as the different shades of red and blue, tag
fluid elements).

In the comoving frame the velocity field  has two 
stagnation points. The stream tube starting at the forward stagnation point
and the stream tube ending at the backward stagnation point
transversely intersect along a finite number of streamlines,
giving rise to a three-dimensional chaotic tangle.
The third part of the video shows the stream tube that emanates from the
forward stagnation point. 
With the time frozen at $t=0$,
this surface is shown from a point of view
that moves first only zonally and then both zonally
and meridionally.

\section*{References}

\parindent 0cm
\parskip 0.2cm

[1] S.P. Thompson 1910 {\em The Life of William Thomson, Baron Kelvin of Largs}, volume~1. McMillan. 

[2] J.J. Thomson 1883. {\em A Treatise on the Motion of Vortex Rings}. McMillan.

[3] W.~Thomson (Lord Kelvin) 1875. Vortex statics. {\em Proceedings of the Royal Society of Edinburgh}, 9:59--73.

[4] O.~Velasco~Fuentes 2010. Chaotic streamlines in the flow of knotted and unknotted vortices.
{\em Theoretical and Computational Fluid Dynamics}, 24:189--193.

\end{document}